\documentclass{article}
\usepackage[T1]{fontenc} % add special characters (e.g., umlaute)
\usepackage[utf8]{inputenc} % set utf-8 as default input encoding
\usepackage{ismir,amsmath,cite,url}
\usepackage{graphicx}
\usepackage{xcolor}
\usepackage{booktabs}
\usepackage{makecell}
\usepackage{enumerate}
\usepackage{dblfloatfix}
\usepackage{amsfonts}
\usepackage{amsmath}
\usepackage{amssymb}

\newcommand{\tb}[1]{\textcolor{blue}{#1}}

\usepackage{lineno}
\title{Tracing Affordance and Item Adoption on Music Streaming Platforms}

\multauthor
{Dougal Shakespeare$^1$ \hspace{1cm} Camille Roth$^1$} {\\
  $^1$ Computational Social Science Team, Centre March Bloch, Berlin\\
{\tt\small dougal.shakespeare@cmb.hu-berlin.de, roth@cmb.hu-berlin.de}
}

\def\authorname{F. Author, S. Author, and T. Author}

\begin{document}
\newcommand{\ul}{\em}
\newcommand{\hidefornow}[1]{}
\maketitle
\begin{abstract}

Popular music streaming platforms offer users a diverse network of content exploration through a triad of affordances: \textit{organic}, \textit{algorithmic} and \textit{editorial} access modes. Whilst offering great potential for discovery, such platform developments also pose the modern user with daily adoption decisions on two fronts: platform affordance adoption and the adoption of recommendations therein. Following a carefully constrained set of Deezer users over a 2-year observation period, our work explores factors driving user behaviour in the broad sense, by differentiating users on the basis of their temporal daily usage, adoption of the main platform affordances, and the ways in which they react to them, especially in terms of recommendation adoption. Diverging from a perspective common in studies on the effects of recommendation, we assume and confirm that users exhibit very diverse behaviours in using and adopting the platform affordances. The resulting complex and quite heteregeneous picture demonstrates that there is no blanket answer for adoption practices of both recommendation features and recommendations.
\end{abstract}

\section{Introduction}\label{sec:introduction}
Today, the modern music streaming platform architecture is a far cry from the digital repository which it once was. Increased diversification at a platform level has resulted in a now common triad of platform affordances, \textit{organic} ($O$), \textit{algorithmic} ($A$) and \textit{editorial} ($E$), which allow users to explore a platform's ever expanding musical catalogue through novel paths.
$A$ affordances refer to the platform's plethora of Recommender System (RS) architectures (\hbox{e.g.,} the popular Flow playlist on Deezer) whilst $E$ affordances correspond to mostly human curated playlists (such as recommended playlists variously called “10s electronic”, “Rock \& Chill”, etc.). We characterise all remaining modes of access under $O$ including for instance the search bar, user-constructed playlists, and more broadly, modes of content access which do not utilise any degree of recommendation.
No longer is it necessary for the avid music listener to spend hours on end organically trawling through digital repositories to find their `niche' but rather, they are free to draw upon $A$ and $E$ affordances to further complement or even replace their exploration in this space. 
Nonetheless, while it is true that affordance diversification yields increased potentials for exploration this comes at the price of a greater emphasis on the user's to utilise the platform - questions of what affordance to utilise and what items therein to adopt %(i.e. transfer into one's organic catalogue) 
quickly become frequent decisions the modern user must face in their daily platform usage. 

In our work we centre our attention on this notion of adoption on two fronts: \emph{affordance adoption} ($O,A,E$) and \emph{item adoptions} therein (\hbox{i.e.} adoption into one's organic catalogue: $A\rightarrow O\: \| \:E\rightarrow O$) through the introduction of the novel metric of adoption. % ($AD$). 
Tracing user listening practices on the popular music streaming platform, Deezer our work sheds light on the varied and often heterogeneous nature of adoption and behavioural differences amongst users.
The contribution of this work sits within the growing body of state of the art literature which appraises the interconnected effects of human behaviour and algorithmic influence through an organic comparison -- a contribution of high relevance to music streaming providers and user's alike to comprehend the bi-directional affect of recommendation upon user preference and downstream platform behaviour. 
%
% TODO: add here structure of document once we fully know...
\section{Literature Review}
\subsection{Appraising Algorithmic Influence via an Organic Comparison}
Whilst historically the primary role of a music Recommender System (mRS) on streaming platforms was to facilitate the efficient personalised exploration of a platform's often vast musical catalogue thereby minimising the risk of \textit{choice overload}, a substantial body of multi-disciplinary literature \cite{youtube, Seaver2019, Bonini, filter-bubbleog} points towards the same conclusion that user exploration nonetheless may remain confined to a minute fraction of homogeneous musical content -- a phenomenon famously denoted as \textit{‘filter bubble’} \cite{filt-bubble}. Similarly, simulation based RS literature has additionally explored the tendency for feedback loops to emerge in a plethora of domains \cite{Ferraro_2020, xavier, Jannach2015WhatRR, zhang}.
Nonetheless, while such practices are clearly outlined in literature, a less trivial second order question still remains ambiguous:
\newline\newline
\textit{To what degree is user platform behaviour primarily a product of algorithmic influence or rather, an autonomous organic process imposed by the user themselves?}
\newline\newline
In light of such questions,  a novel branch of the state of art has sought to measure algorithmic influence by drawing parallels to a user's organic platform behaviour that is, without algorithmic influence. Roth conceptualises this debate through what he coins the \textit{‘ROM-COM dichotomy’} \cite{Roth2019} -- the tendency for filtering algorithms to either Read Our Minds (ROM) acting as cognitive aids to facilitate organic exploration or rather, Change Our Minds (COM) algorithmically distorting a prior organic preference.  

Literature appraising algorithmic influence through an organic reference has been applied to a range of multi-disciplinary contexts. In the works of Bakshy et al.~\cite{Bakshy2016}, the potential for Facebook's NewsFeed algorithm to expose users to cross-partisan content is appraised by drawing parallels to user organic explicit preference and subsequent item consumption. Their work finds filtering algorithms to have minimal effect in reducing cross-partisan information in comparison to the organic selection processes imposed by the user suggesting a user's diversity limitation may be principally due to human pre- and post-selection.
In the music domain, Epps-Darling et al. \cite{darling} study the role of Spotify’s algorithmic influence on gender representation through an organic reference point. Their findings show user’s organic preference towards male artists to be marginally stronger than that recommended by Spotify’s mRS, again suggesting the human organic bias to be stronger than that generated algorithmically. 
Anderson et al. \cite{Anderson2020} also analyse diversity with respect to organic vs programmed (algorithmic/editorial) listening events on Spotify. Through the generation of a usage-based embedding space, they find algorithmically-driven exploration in this space to be less diverse than organic, providing evidence for a COM effect. 
\hidefornow{More closely related to this work, Villermet et al. \cite{Villermet2021} appraise the role of algorithmic influence via an organic comparison on Deezer. Their findings show distinct behavioural difference between user vary in the degree of affordance adoption both in terms of activity and the item content they consume - users who favour algorithmic affordances consume less mainstream content which leads to the introduction of what they coin, a \textit{filter-niche} as opposed to Pariser’s popularised notion of the \textit{filter bubble} \cite{filt-bubble}.}
\newline
\newline
Overall, current literature paints a picture of divergent music streaming platform practices dependent upon the user's degree of organic-algorithmic usage and actualised with respect to context. Thus, we commence our work with the prior hypothesis that platform use behaviour is largely varied - there is no \textit{average user} for which a blanket answer to algorithmic influence may be applied.
\subsection{Item Adoption in Recommender System}
Whilst item adoption in terms of consumption confined to a given affordance has been covered extensively and critically both in terms of user studies \cite{Ekstrand2018,Jin} and user modelling \cite{pichl, aridor2020, schedl}, literature concerning item adoption in relation to the dynamic transfer of items across affordance (e.g. from an external affordance into ones organic catalog) remains to this date, sparse. 
Nonetheless, at a higher level, the adoption of music streaming platforms as such, independent of the affordances they offer, has been notably studied in the field of Cultural Studies. In the works of Datta et al. \cite{Datta}, streaming adoption is shown to lead to substantial increases in both the quantity and diversity of music consumed by a user. Similarly, Rushan et al. \cite{Rushan} also study factors of the platform interface which in itself, determine a consumer's decision to adopt music streaming platforms as a result of increased platform familiarisation.
Still, adoption with respect to transitions across platform affordances remains a literature void for which this work seeks to fill. 
\subsection{Temporal Dimensions of mRS Usage}
Time of day information has been evidenced to be an important signal in disentangling platform behaviour \cite{rockingclock} and thus, has in recent years become a signal commonly utilised in context dependent RS literature \cite{Baltrunas2009, Sergey, tod_music} to provide increased personalisation and accuracy. What is more, user studies have also revealed that both the time of day and week can play a substantial role in mediating user platform experience and downstream projected user behaviour \cite{krause, HeadphonesOT}. Utilising such rich temporal signals encapsulated within listening logs, our work seeks to explore the degree to which adoption on both fronts may differ across temporal daily usages.
\section{Methodology}\label{sec:page_size}
\subsection{Listening Events Data Set}
\begin{figure}[t]
 \centerline{
 \includegraphics[width=\columnwidth]{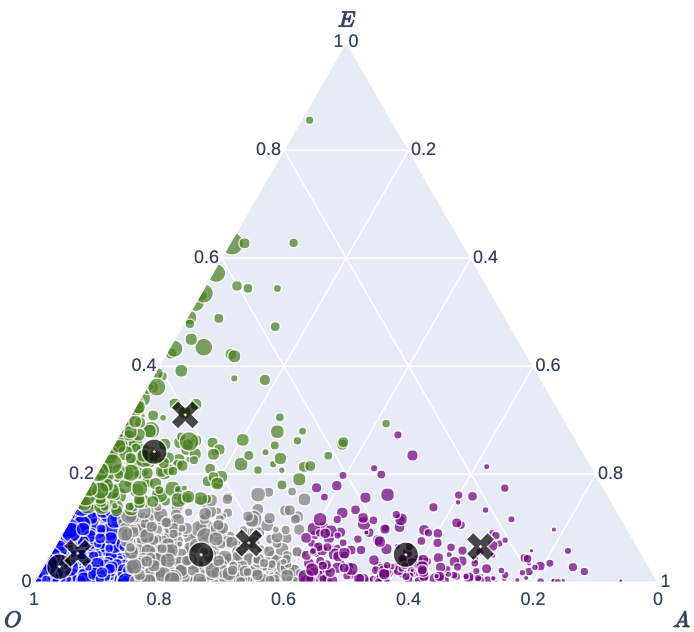}}
 \caption{Ternary plot of affordance adoption classes \o+ (blue), o (grey), e (green), a (purple), disks represent centroid positions. Crosses represent corrected centroid positions after taking into account the pre-adoption origin of plays (see Sec.~\ref{sec:discussion}).}
 \label{fig:affordClust}
\end{figure}
%
%We work with about 2 years of listening histories from a cohort of 2701 Deezer users who all register within a 1 month period (September 2017) and remain active in their use of the platform over the entire observation period. To impose a formal constraint of \textit{‘activeness’} we define a maximum inter-event time threshold of 10 days thus ensuring that users rely on the platform for a regular source of music.
We work with about 2 years of listening histories from about 13k Deezer users who registered within a 1 month period in September 2017. We further focus on users who remained active %in their use of the platform 
over the entire observation period: formally, we impose a maximum inter-event time threshold of 10 days thus ensuring that users rely on the platform for a regular source of music. This eventually yields 2701 users.

We discard listening events <30s\hidefornow{ as per \cite{Villermet2021}} as these are deemed as so-called \textit{‘skips’}. We further merge unique song identifiers which share identical audio embeddings in a pre-build latent space supplied by Deezer (see \cite{metricl} for use case / generation details). This prevents double couting of identical songs which may have been mis-labeled as distinct. %are ultimately considered identical thereby preventing double counting in subsequent computations. 
For each listening event we characterise the mode of access used to retrieve the content of which on Deezer there exists a triad of affordances: \textit{organic} ($O$), \textit{algorithmic} ($A$) and \textit{editorial} ($E$).
{\subsection{Defining user classes}}

{\textbf{Affordance Adoption.}}
%We explore two dimensions of platform behaviour for which prior literature has suggested play a crucial role %in mediating algorithmic influence - affordance and time of day preference. In the following section we %detail the methodology surrounding static and temporal clusters which act as pre-requisites for the latter %behavioural analysis described.
%
We capture user adoption of platform affordances through the proportion of content accessed via each of the platform's three main affordances after aggregating listening histories. For a given user, their listening history is represented by the temporally-ordered list of listening events over our observation period, formally defined as:
%
%$$LH = \{s_1, s_2, ... s_n\}$$
%
{$$P=\big((s_i,t_i,f_i)\big)_{i\leq n}$$
where $s_i$ is the song ID, $t_i$ the timestamp, and $f_i\in\{A,E,O\}$ the affordance used for the $i$-th listening event. We accordingly denote the sublist of $P$ restricted to a given affordance $F$  as $P_F=\big((s_i,t_i,f_i)\big)_{i\leq n\,\wedge\, f_i = F}$}
%\tg{Let $t(s)$ represents the list of all time stamps in which song $s$ was played by a user sequentially. Similarly, let $O,A,E \subset LH$ and $(t_o,t_a,t_e)$ denote the ordered list of a user's \textit{organic, algorithmic} and \textit{editorial} plays and timestamps respectively.}
%
{Considering the proportion of plays accessed \textit{organically, algorithmically} and \textit{editorially} respectively, the affordance profile of a given user is defined by the triplet $(|P_O|/|P|, |P_A|/|P|, |P_E|/|P|)$ which sums to 1 and} can be represented as a barycentric coordinate in ternary space (see Figure \ref{fig:affordClust}). Performing a k-means clustering ($k = 4$) across affordance profiles yields 4 distinct classes of users which we label as follows: very organic `$o+$' (1786 / 65.98\%), organic `$o$'  (429 / 15.85\%), algorithmic `$a$' (224 / 8.27\%), organic/editorial `\oe' (268 / 9.90\%). 
We note that we rather deal with bins, areas or classes than with well-separated clusters per se. Thus, from herein we refer to affordance adoption clusters as classes. 
Already, %we start to see 
a highly varied picture of affordance adoption on the platform emerges\hidefornow{ coherent with the cohort of Deezer users analysed in \cite{Villermet2021}} along with %. Whilst users adopt platform affordances to various degrees we identify 
a shared preliminary benchmark: for all classes, users on average display some degree of $O$ adoption whilst the same cannot be said for $A$ and $E$.
Indeed, at first sight it appears Deezer users do not typically adopt affordances on all fronts but rather use the platform predominately as an organic catalogue much the way one would search through a traditional song library, albeit a much larger one here. However, as we shall later detail, a user's tendency to consume mostly $O$ content may be misleading: if a significant proportion of a user's $O$ catalogue is a product of $A$ or $E$ adoption the very definition of what it means to adopt $O$ affordances is brought into question.

\medskip\noindent\textbf{Exploration Behaviour.}
Beyond sheer user activity over the entire observation period denoted by play counts $|P|$, we consider % 2 behavioural measures denoting user activity: (1) \textbf{activity} ($A$) quantified for each user as the number of plays, $P$ over the entirety of our observation period.
%\newline\newline
%\noindent (2) 
a notion of \textit{redundancy} \cite{HeadphonesOT} quantified as a measure of how much a given user saturates their listening catalogue(\hbox{i.e.,} plays the same songs repeatedly), formally defined as $R=1-|S|/|P|$ where $S=\{s|(s,t,f)\in P\}$ is the set of unique songs in $P$. % transform of list $P$ and denotes the unique  $$R = 1 - S/P$$where $S = |LH|$.

%To capture the influence of affordance adoption and temporal daily adoption differences w
We additionally characterise the diversity of a user's exploration %at a low audio level that is, within 
using a pre-built 32 dimensional latent space $\mathcal{E}$ constructed from low-level audio features via metric learning \cite{metricl}. The audio embeddings %which inhabit this space 
were primarily used by Deezer for the task of artist disambiguation (\hbox{i.e.,} where artists had the same name but were stylistically unique). Thus, the construction of $\mathcal{E}$ %such embeddings 
%ensures that distance is 
maximises %for 
acoustically dissimilar artists while acoustically similar artists remain close in this space. For each user, %We utilise such embeddings to assess a user's diverse exploration at the low audio level, computing for each user 
we compute their average pairwise \emph{cosine distances} between audio embeddings $\mathcal{E}_s$ for each $s \in S$ and ultimately, report average values.

%
%Before tackling this pressing questions, we now consider the effect of time of day upon affordance adoption.
%
\begin{figure}[t]
 \centerline{
 \includegraphics[width=\columnwidth]{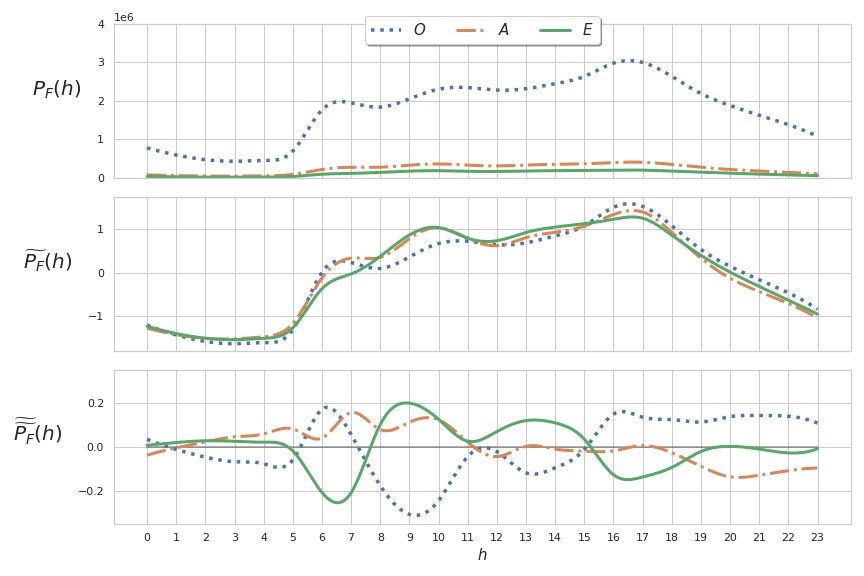}}
 \caption{Normalised platform affordance time-of-day activity. Aggregate levels are shown (top) followed by z-normalised activity (middle) and residual de-trended activity levels (bottom).}
 \label{fig:affordPlatform}
\end{figure}
%
%\subsection

\newcommand{\PP}{\mathcal{P}}

\bigskip\noindent\textbf{Temporal Time of Day Analysis.}
Usage also varies significantly over the elapsed day. We consider platform activity with respect to \emph{time-of-day} at an aggregate level across \hidefornow{$O,A,E$}all affordances (Fig.~\ref{fig:affordPlatform}, top row). % ($tod$).
%Detailing first aggregate absolute levels (see 
We observe that it predominantly consists of $O$ followed by $A$ and $E$ at much lower magnitudes, consistent with what has been previously detailed. {Without loss of generality, we denote $\PP$ as the platform-level history for all users (\hbox{i.e.,} the whole dataset).} %This comes as no surprise since the majority of Deezer users analysed are both $o+$ and as previously detailed, all remaining affordance taxonomies consume some degree of $O$ content. 
{To analyse temporal trends independent of magnitude,
we consider hourly play counts for each affordance  $\PP_F(h)=\left|\big((s_i,t_i,f_i)\big)_{i\leq n\,\wedge\,(t_i\text{'s hour}=h)\,\wedge\,(f_i=F)}\right|$, to which 
%thus providing a clearer insight into the Deezer platform's daily affordance access trends 
we subsequently apply a z-normalisation relative to daily play count averages
%The normalization mathematically 
defined as:
%
%$$pc^{\text {norm }}(h)=\frac{pc^{a b s}(h)-\mu_{pc^{a b s}}}{\sigma_{pc^{a b s}}}$$
%
$$\widetilde{\PP_F}(h)=\frac{\PP_F(h)-\big\langle \PP_F(h)\big\rangle_{h}}{\sigma_{\PP_F(h)}}$$}%
%where $pc(h)$ denotes aggregate playcounts at hour, $h$. 
These normalised activity levels (Fig.~\ref{fig:affordPlatform}, second row) reveal three peaks of gradually increasing magnitude, respectively in the early morning, morning, and afternoon (16-17:00), % which we label as follows: \textit{‘early rise’, ‘wake up’, ‘afternoon’} based upon their temporal location. As a rule of thumb, it appears platform activity for each access mode gradually increases throughout the day until the ultimate aforementioned \textit{‘afternoon’} peak at approximately
from which a gradual decay in activity is experienced. %We hypothesise this may imply platform activity peaks when users are able to devote their full attention (i.e. before or prior to work) however this should be verified by further user studies.
To capture temporal adoption variations of affordances with respect to one another, we finally % varies temporally at a platform level we next 
apply so-called \textit{‘detrending’} as per \cite{toole} by comparing {the above z-normalisations $\widetilde{P_F}(h)$ relative to other affordances, formally defined as:
$$
%pc^{\text {res }}(h)=pc^{\text {norm }}(h)-\bar{pc}^{\text {norm }}(h)
\widetilde{\widetilde{\PP_F}}(h)=\widetilde{\PP_F}(h)-\big\langle\widetilde{\PP_F}(h)\big\rangle_F
$$}%
%
%capture platform activity for each access mode with respect to average normalised levels at each time step, $h$. This 
%which yields a more clear representation of relative structural trend differences which reside within affordance activity. 
%distinct platform affordance trends now emerge. For the early morning peak, we observe the utilisation of $O$ to be more prominent (independent of magnitude) whilst there is minimal tendency for $E$ affordances to be adopted respective to $A$ and $E$ affordance life cycles in this time block. The $E$ activity appears to be more bi-modal, suggesting user's tendency to favour $E$ access in the early hours of the morning is weaker than that of $O$ and $A$ affordances. 
%
%Between the early morning and morning peaks the platform trend for $A$ and $E$ affordances to be adopted is far greater than that of $O$ - it is only at the ultimate afternoon peak that we observe a return of the $O$ residual peak however this quickly decays into the early hours of the morning to be replaced by $A$ and $E$ activity. 
%
%From this preliminary analysis affordance adoption at a platform level is clearly shown to vary temporally across hours of the day. Platform affordance adoption cycles of $O$ access is found to be favoured in the early morning and afternoon hours of the day whilst the intermediate hours are filled with a period of $A$ and $E$ activity. 
Affordance adoption at a platform level clearly varies across hours of the day (Fig.~\ref{fig:affordPlatform}, bottom row). For one, in the early morning and from the afternoon hours and on into the evening, we observe the tendency to favour $O$ to be more prominent, respective to $A$ and $E$ affordance life cycle in the same time blocks (\hbox{i.e.} independent of magnitude). Otherwise either $A$ or $E$ seem to be favoured (again, in relative trend) and essentially appear to exhibit bi-modal relative adoption peaks across the day, albeit at different moments (rather in the morning for $A$ and in the early morning and early afternoon for $E$).

%whilst there is minimal tendency for $E$ affordances to be adopted respective to $A$ and $E$ affordance life cycles in this time block. 
 
\begin{figure}[t]
 \centerline{
 \includegraphics[width=\columnwidth]{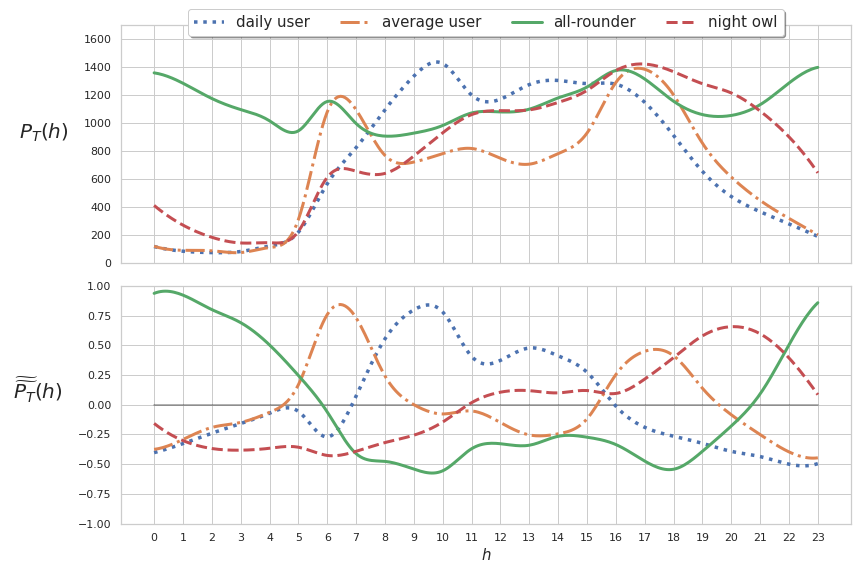}}
 \caption{Normalised time-of-day activity levels for each time-of-day class. Aggregate levels are shown (top) followed by residual de-trended activity levels (bottom).}
 \label{fig:todClust}
\end{figure}
%
%Having focused on adoption at a platform level, 
We complement the temporal platform-level analysis by examining daily patterns at the user level, as %. The value of the complementary user-centric analysis lies in the fact that 
users will commonly have varied activity levels on music streaming platforms \cite{Mehrotra2020}. %and thus, considering aggregated play counts that fall within $t(h) < t(h+1)$ may fail to miss the user relative semantics of activity. 
{To this end we consider user-level hourly activity aggregated over all affordances %$P_F(h)$ and thus 
$P(h)=\left|\big((s_i,t_i,f_i)\big)_{i\leq n\,\wedge\,(t_i\text{'s hour}=h)}\right|$.
%Following a similar notation to \cite{HeadphonesOT}, let $p(h_u)$ denote the proportion of plays for user $u$ which occur between hours $h$ and $h+1$ aggregated over the entirety of our observation period. For each $u$ we construct a 24-values vector $(p(0{_u}), . . . , p(23_{u}))$ for 
As in \cite{HeadphonesOT}, we subsequently cluster users using k-means ($k=4$ again) applied on $P(h)$ as a 24-dimensional unit vector}. We observe 4 distinct user-centric behavioural dynamics which we label as follows: the \textit{`average'} user (726 / 26.82\%), the \textit{‘night owl’} (928 / 34.28\%), the \textit{`all rounder’} (245 / 09.05\%) and the \textit{‘daily’} user (808 / 29.85\%). 
{We represent the variations of $P(h)$ on Fig.~\ref{fig:todClust} by applying the same type of normalization as used above for platform-level quantities $\PP(h)$, except that we consider temporal clusters $T$ instead of affordances $F$ \hbox{i.e.,} $\PP_T(h)$ in place of $\PP_F(h)$.}
%

%
%
%
%
%Nicheness: defined at a high level to be the inverse %of popularity. To quantify nicheness we bin artists %into quartiles based upon aggregated play counts. We %thus label our 4 levels of popularity as follows: %(super stars, top head, long tail, distant tail).
%
%\begin{figure}
%     \centerline{
%     \includegraphics[width=1\columnwidth]{2021/latex/figs/affordeg2.png}}
%     \caption{Example sequential listening histories (right) for a given song $s$ %which could belong to each Adoption set (left, bottom down): \textit{Adoption %Feasible}, \textit{Adoption Unfeasible} and \textit{Adoption Achieved}.}
%\label{adoptFig}
%\end{figure}
\subsection{Item Adoption}
To measure the tendency for users to adopt algorithmic or editorial songs into their organic catalogue we define the novel measure of \textit{adoption} {$\alpha$}. 
%
%For both algorithmic and editorial affordances, the means of computing adoption are identical and thus, we use the dummy notation \tb{$F$} to denote either the \tb{affordances} $A$ or $E$.  
%
At a high level, an item can be said to be adopted the first time it has been played organically by a user given that the song was first recommended {through some affordance $F$} and not played organically as a prior. {Since we are interested in item adoption and thus unique songs, we now focus on song sets rather than lists of plays \hbox{i.e.,} $S$% and, respectively for each affordance, $S_A$, $S_E$ and $S_O$%
. We} outline two possible mutually exclusive song sets, denoted $\phi$ and $\rho$, %which a song will belong to thereby 
to differentiate songs which could have been adopted yet were not, from the ones which were actually adopted: 
\newline
\textbf{-- Adoption feasible, {$\phi$}}, denoting the set of songs which were played through {$F$} but not via $O$ and thus, had the potential to be adopted yet were not:
{\[\phi_F = 
%S_F \setminus S_O
\left\{ s\in S \:\Big|\:
\exists (s,t,F)\in P,
\nexists (s,t',O)\in P
%i, s_i=s, f_i=F \wedge \nexists i, s_i=s, f_i=O
\right\}
\]}%
%
% \textbf{S2: Adoption unfeasible \tb{($\upsilon$)}:} set of songs which were \tb{played via $O$ prior to an $F$ play, if any.} %either never played via $C$ or were via $O$ prior to any $C$ play. 
% %
% \tb{$$\upsilon_F = \left\{ s \Big| \exists i, s_i=s, f_i=O, \nexists j, s_j=s, f_j=F, t_j<t_i\right\}%
% $$}
% % \begin{equation*}
% %   \begin{array}{l}
% %     AUF = \{s : s \in (C \cup O) / C \: \vee \\
% %     \:\:\:\:\:\:\:\:\:\:\:\:\:\:\:\:\:\:\:\:\:\:\:\:s \in O \cap C \wedge \min(t_o(s)) > \min(t_c(s))\}
% %   \end{array}
% % \end{equation*}
%
{\bf-- Adoption {realised, $\rho$}}, denoting the set of songs which were played via {$F$} as a prior and were consumed through $O$ at least once subsequently: 
{%$$\rho_F=\left\{ s \:\Big|\: \forall t_i |\substack{s_i=s\\f_i=O},\: \exists t_j |\substack{s_j=s\\f_j=F}, t_j<t_i\right\}$$%
\[%
\rho_F=\left\{ s \in S \:\Big|\: \exists (s,t,F)\in P,
t < \min_{(s,t',O)\in P}t'  \right\} 
\]}%
%$$AA = \{s_i : s_i \in O \cap C \wedge \min(t_o(s_i)) < \min(t_c(s_i))\}$$
%
%With \tb{$\phi$ and $\rho$ we may compare the number of songs which could have been adopted from the ones which were. 
Furthermore, if a user is exposed to more recommendations before ultimately making the decision to adopt, this may indicate a weaker influence of the platform's affordance or that adoption is less likely to be a direct product of it (for instance the user might be more likely to have heard the song from an external source such as the radio). To capture this intuition, we introduce $r_F(s)$, the number of recommendations of song $s$ which appeared through $F$ before organic adoption:
\[r_F(s) = \left| \left\{ (s,t,F)\in P \:\big|\: t < \min_{(s,t',O)\in P}t'  \right\} \right|\]
to which we apply a polynomial scaling function which decays to give more weight to lower numbers of recommendations %a higher weighting to smaller differences in comparison to those of a larger magnitude 
- a similar practice to how listening counts are often scaled logarithmically in mRS literature \cite{Dean_2020, Jawaheer}. 

We assess the relative impact of item adoption at two levels of abstraction. Foremostly, with respect to the {number} of items which both could have been and were adopted by the user {through $F$ \hbox{i.e.,} $|\phi_F| + |\rho_F|$}. Formally let this be defined by:
% we apply a polynomial scaling to the number of recommendations of song $s$ which appeared before %the ultimate
% organic adoption. The scaling function decays to apply a higher weighting to smaller differences in comparison to those of a larger magnitude - a similar practice to how listening counts are often scaled logarithmically in mRS literature \cite{Dean_2020, Jawaheer}. We assess the relative impact of item adoption at two levels of abstraction. Foremostly, with respect to the proportion of items which both could have been and were adopted by the user ($|AA| + |AF|$). Formally let this be defined by:
%
%$$AD_{ca} = \dfrac{\sum_{s \in AA} \dfrac{1}{|t_c(s) < \min{t_o(s)}|}^\alpha}{|AA|+|AF|}$$
{\[
\alpha_{F}
=\dfrac{\sum_{s\in\rho}{r_F(s)}^{-\lambda}}{|\phi_F|+|\rho_F|}
\]}%
where {$\lambda \in (0,1]$} is a hyperparameter which affects the degree of polynomial decay with respect to algorithmic impact. In our experiments we set the value of {$\lambda = 0.5$}. We note our choice of {$\lambda$} is cautious and should in future work be more refined with statistical and qualitative user studies exploring the role of repeated affordance recommendation prior to adoption. 

Secondly, we normalise adoption with respect to the number of unique items consumed via $O$, thereby capturing the relative impact of algorithmic adoption in a user's overall organic listening catalogue. Formally,
%
%$$AD_{co} = \dfrac{AD_c}{|O|}$$
\[\alpha'_F=|\rho_F|/|\{ s\in S \:|\:\exists(s,t,O)\in P\}|\]%
We note that this value is bounded by the number of organic streams in a user's listening history but nonetheless, we deem this to be a useful measure to capture the influence of item adoption in bringing into question the very meaning of what is deemed organic.
\begin{table*}[t!]
\centering
\small\addtolength{\tabcolsep}{-2pt}
  \begin{tabular}{c>{\hspace{.3em}}cccccp{1em}ccccc}
    \toprule
    &\multicolumn{5}{c}{o+}&&\multicolumn{5}{c}{o}\\\cmidrule{2-6}\cmidrule{8-12}
    &\textit{ar}&\textit{au}&\textit{no}&\textit{du}&\textit{all, (znorm)}&
    &\textit{ar}&\textit{au}&\textit{no}&\textit{du}&\textit{all, (znorm)}\\
    \cmidrule{1-6}\cmidrule{8-12}
    \makecell{$|P|$\\} &\makecell{\textbf{*26.62K}}&\makecell{15.59K}&\makecell{*20.87K}&\makecell{15.94K}&\makecell{18.58K, (0.21)}&
    &\makecell{\textbf{26.09K}}&\makecell{15.63K}&\makecell{19.68K}&\makecell{16.88K}&\makecell{18.01K, (0.18)}\\
    %\hline
    \makecell{$R$} &\makecell{\textbf{*0.90}}&\makecell{0.85}&\makecell{0.86}&\makecell{*0.83}&\makecell{0.85, (0.02)}&
    &\makecell{\textbf{*0.82}}&\makecell{0.77}&\makecell{0.77}&\makecell{0.77}&\makecell{0.77, (0.02)}\\
    %\hline 
    \makecell{$\alpha_{A}$} & \makecell{\textbf{0.28}}&\makecell{0.25}&\makecell{0.28}&\makecell{0.25}&\makecell{0.25, (0.06)}&
    &\makecell{\textbf{0.15}}&\makecell{0.14}&\makecell{0.15}&\makecell{0.14}&\makecell{0.14, (0.07)}\\
    %\hline
    \makecell{$\alpha_{E}$} & \makecell{0.26}&\makecell{\textbf{0.27}}&\makecell{0.24}&\makecell{*0.21}&\makecell{0.25, (0.10)}&
    &\makecell{0.18}&\makecell{\textbf{0.19}}&\makecell{0.16}&\makecell{0.15}&\makecell{0.17, (0.08)}\\
    %\hline
    \makecell{$\alpha'_{A}$} &\makecell{0.01}&\makecell{0.02}&\makecell{\textbf{0.02}}&\makecell{0.02}&\makecell{0.01, (0.05)}&
    &\makecell{0.06}&\makecell{\textbf{0.06}}&\makecell{0.06}&\makecell{0.05}&\makecell{0.06, (0.03)}\\ 
    %\hline
    \makecell{$\alpha'_{E}$} &\makecell{0.02}&\makecell{\textbf{0.02}}&\makecell{0.02}&\makecell{0.02}&\makecell{0.02, (0.08)}&
    &\makecell{\textbf{0.03}}&\makecell{0.02}&\makecell{0.02}&\makecell{0.03}&\makecell{0.02, (0.04)}\\ 
    %\hline 
    \makecell{$\mathcal{E}$ dist.} & \makecell{0.28}&\makecell{0.29}&\makecell{0.29}&\makecell{\textbf{0.29}}&\makecell{0.28, (0.02)}&
    &\makecell{0.28}&\makecell{0.29}&\makecell{\textbf{0.30}}&\makecell{0.29}&\makecell{0.29, (0.01)}\\
    \\
    &\multicolumn{5}{c}{a}&&\multicolumn{5}{c}{\oe}\\
    \cmidrule{1-6}\cmidrule{8-12}
    \makecell{$|P|$\\} &\makecell{\textbf{41.20K}}&\makecell{15.06K}&\makecell{16.18K}&\makecell{19.75K}&\makecell{19.01K, (0.40)}  &
    &\makecell{\textbf{23.80K}}&\makecell{15.40K}&\makecell{19.68K}&\makecell{23.09K}&\makecell{20.14K, (0.15)}\\
    %\hline
    \makecell{$R$} &\makecell{\textbf{*0.85}}&\makecell{0.77}&\makecell{\textit{0.74}}&\makecell{0.77}&\makecell{0.77, (0.04)}&
    &\makecell{0.77}&\makecell{0.74}&\makecell{\textbf{0.77}}&\makecell{0.76}&\makecell{0.76, (0.01)}\\
    %\hline
    \makecell{$\alpha_{A}$} & \makecell{\textbf{0.10}}&\makecell{0.09}&\makecell{0.09}&\makecell{0.09}&\makecell{0.09, (0.04)}&
    &\makecell{0.17}&\makecell{0.16}&\makecell{\textbf{0.20}}&\makecell{0.15}&\makecell{0.17, (0.08)}\\
    %\hline 
    \makecell{$\alpha_{E}$} & \makecell{0.20}&\makecell{0.14}&\makecell{0.15}&\makecell{0.13}&\makecell{0.14, (0.13)}&
    &\makecell{0.14}&\makecell{0.14}&\makecell{\textbf{0.16}}&\makecell{0.13}&\makecell{0.14, (0.08)}\\
    %\hline 
    \makecell{$\alpha'_{A}$} &\makecell{0.08}&\makecell{\textbf{0.08}}&\makecell{0.08}&\makecell{0.08}&\makecell{0.08, (0.02)}&
    &\makecell{\textbf{0.03}}&\makecell{0.02}&\makecell{0.02}&\makecell{0.02}&\makecell{\textbf{0.02}, (0.14)}\\ 
    %\hline
    \makecell{$\alpha'_{E}$} &\makecell{0.03}&\makecell{\textbf{0.03}}&\makecell{0.02}&\makecell{0.03}&\makecell{0.03, (0.04)}&
    &\makecell{0.05}&\makecell{0.05}&\makecell{\textbf{0.05}}&\makecell{0.05}&\makecell{\textbf{0.05}, (0.04)}\\ 
    %\hline
    \makecell{$\mathcal{E}$ dist.} & \makecell{0.29}&\makecell{0.30}&\makecell{0.29}&\makecell{\textbf{0.30}}&\makecell{0.30, (0.01)}&
    &\makecell{0.28}&\makecell{0.29}&\makecell{0.28}&\makecell{\textbf{0.29}}&\makecell{0.29, (0.01)}\\
    \bottomrule
  \end{tabular}
  \caption{Experimental results across two static affordance and temporal \textit{time of day} user classes. Values in bold represent the top value, while marked with * are results where the difference is statistically
significant (two tailed t-test, $\alpha=0.05/n$ after Bonferroni correction).}.
  \label{tab:recap}
\end{table*}%

\section{Results}
\begin{figure}[t]
 \centerline{
 \includegraphics[width=1\columnwidth]{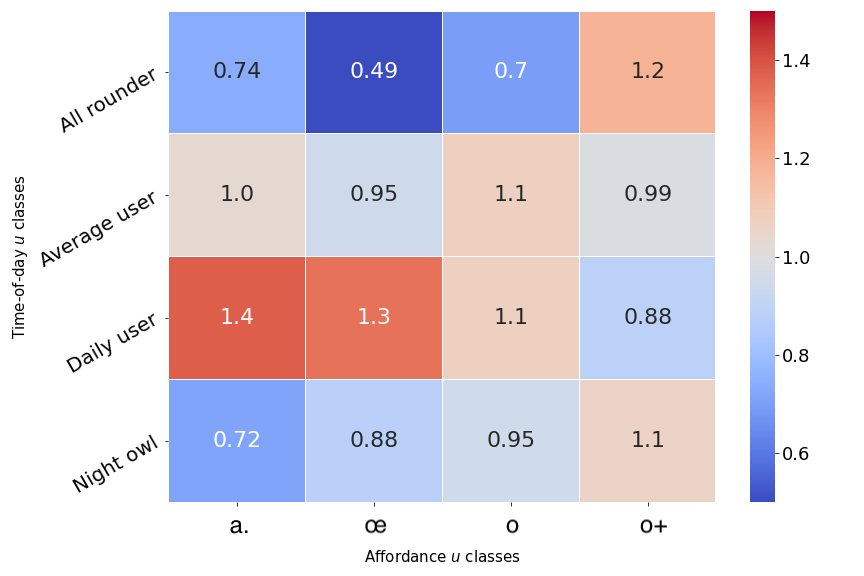}}
 \caption{Affordance \hbox{vs.} time-of-day distributions. 
 %Values of 1 indicate play counts equivalent to that found globally in the dataset whereas values above and below 1 represent over and under representation respectively.
 Values are normalised such that above or below 1 indicate respectively similar, over- or under- representation of affordance classes respective to those found globally.
 }
\label{fig:affordTodDist}
\end{figure}

\textbf{Temporal Affordance Adoption Variations.}
We first examine the distribution of affordance classes across {time-of-day} classes. As shown in Figure~\ref{fig:affordTodDist} we observe two fundamental preliminary findings: (1) daily users are more heavily composed of both $a$ and \oe users respective to other {time-of-day} classes. (2) $o+$ users are more proportionally likely to reside within the all rounder and, to a lesser extent, night owl class.
Framed differently, users who adopt almost solely $O$ are more likely to favour platform activity in the evening hours of the day whilst users who more heavily $A/E$-adopt are more likely to favour activity in the day time hours. Once again, our findings reiterate what was observed from our temporal platform evaluation -- %users do not favour affordances equally at all times of the day. 
{the use of recommendation affordances corresponds to different categories of temporal use as well as, we contend, different types of users.}

\smallskip
\noindent\textbf{Characterising platform behaviour.}
To further disentangle the respective use cases we now characterise behavioural dynamics for each time-of-day and affordance class.  Focusing first on 
%Examining first behavioural dynamics of 
affordances, we attain results that go against the grain {of %the traditional 
a} diversity-constraining narrative (see Table \ref{tab:recap}). Users who $A$-adopt more frequently are found to have more diverse exploration in $\mathcal{E}$ whilst maintaining relatively low {redundancy levels as measured by $R$}. It appears the consumption of $A$ content in fact diversifies a user's $P$ at both a behavioural and deeper content-based level whilst on the contrary, $o+$ users are found to saturate their listening catalogue reflected in the high $R$ levels attained.
With regard to item adoption within affordance classes we observe both $a$ and \oe users to have low levels of {$\alpha_{A}$} and {$\alpha_{E}$} respectively. This can be interpreted as a passivity to recommendations - such users are more likely to use $A$ and $E$ affordances regularly but on a so-called \textit{auto-pilot} akin to radio consumption. Nonetheless, when $a$ and \oe users do take the decision to adopt this makes a substantial impact to their $O$ catalogue and thus, the dispersion of users in the $A,E,O$ ternary space as we shall later detail.
Drawing parallels to a more pure organic behaviour through the $o+$, we observe polar opposite dynamics in comparison to the $a$ and \oe users. Whilst $o+$ users $A/E$-adopt sparsely, their ultimate downstream platform use is less recommendation-skeptic reflected in the much higher levels of adoption rates for $A$ and $E$ affordances attained ({$\alpha_{A}= 0.25$, $\alpha_{E}=0.25$}) but with minimal impact to the constitution of their overall $O$ catalogue.
From the detailed findings it is clear that our preliminary assumption that users display varied behavioural dynamics holds true: users diverge dramatically in both their affordance adoption and adoption of items therein and perhaps most importantly, with varying degrees of impact to their overall $O$ catalogue.
%
% HIDDEN 
\hidefornow{
\begin{table*}[t]
\centering
\small\addtolength{\tabcolsep}{-2pt}
  \begin{tabular}{c>{\hspace{.3em}}cccccp{1em}ccccc}
    \toprule
    &\multicolumn{5}{c}{All rounder}&&\multicolumn{5}{c}{Average User}\\\cmidrule{2-6}\cmidrule{8-12}
    &\textit{o+}&\textit{o}&\textit{a}&\textit{\oe}&\textit{all}&
    &\textit{o+}&\textit{o}&\textit{a}&\textit{\oe}&\textit{all}\\
    \cmidrule{1-6}\cmidrule{8-12}
    \makecell{$|P|$\\} &\makecell{\textbf{*26.62K}}&\makecell{\textbf{26.09K}}&\makecell{\textbf{41.20K}}&\makecell{\textbf{23.80K}}&\makecell{\textbf{27.32K}}&
    &\makecell{\tb{\textit{*15.59K}}}&\makecell{\textit{15.63K}}&\makecell{\textit{15.06K}}&\makecell{\textit{15.40K}}&\makecell{15.53K}\\
    %\hline
    \makecell{$R$} &\makecell{\textbf{*0.90}}&\makecell{\textbf{*0.82}}&\makecell{\textbf{*0.85}}&\makecell{0.77}&\makecell{\textbf{0.88}}&
    &\makecell{0.85}&\makecell{0.77}&\makecell{0.75}&\makecell{\textbf{0.77}}&\makecell{0.82}\\
    %\hline 
    \makecell{$\alpha_{A}$} & \makecell{\textbf{0.28}}&\makecell{\textbf{0.15}}&\makecell{\textbf{0.10}}&\makecell{0.17}&\makecell{\textbf{0.25}}&
    &\makecell{0.25}&\makecell{0.14}&\makecell{0.09}&\makecell{\textbf{0.16}}&\makecell{0.25}\\
    %\hline
    \makecell{$\alpha_{E}$} & \makecell{0.26}&\makecell{0.18}&\makecell{\textbf{0.20}}&\makecell{0.14}&\makecell{\textbf{0.25}}&
    &\makecell{\textbf{0.27}}&\makecell{\textbf{0.19}}&\makecell{0.15}&\makecell{\textbf{0.16}}&\makecell{0.21}\\
    %\hline
    \makecell{$\alpha'_{A}$} &\makecell{0.01}&\makecell{0.06}&\makecell{0.08}&\makecell{\textbf{0.03}}&\makecell{0.02}&
    &\makecell{\textbf{0.02}}&\makecell{0.06}&\makecell{0.08}&\makecell{0.02}&\makecell{0.02}\\ 
    %\hline
    \makecell{$\alpha'_{E}$} &\makecell{0.02}&\makecell{\textbf{0.03}}&\makecell{0.03}&\makecell{0.05}&\makecell{0.02}&
    &\makecell{0.02}&\makecell{0.02}&\makecell{0.02}&\makecell{\textbf{0.05}}&\makecell{0.03}\\ 
    %\hline 
    \makecell{$\mathcal{E}$ dist.} & \makecell{0.28}&\makecell{0.29}&\makecell{0.29}&\makecell{0.28}&\makecell{0.28}&
    &\makecell{0.28}&\makecell{0.29}&\makecell{0.30}&\makecell{0.29}&\makecell{\textbf{0.29}}\\
    \\
    &\multicolumn{5}{c}{Night Owl}&&\multicolumn{5}{c}{Daily user}\\
    \cmidrule{1-6}\cmidrule{8-12}
    \makecell{$|P|$\\} &\makecell{20.87K}&\makecell{ 19.68K}&\makecell{16.18K}&\makecell{19.68K}&\makecell{19.90K}  &
    &\makecell{15.94K}&\makecell{16.88K}&\makecell{19.75K}&\makecell{23.09K}&\makecell{17.48K}\\
    %\hline
    \makecell{$R$} &\makecell{0.86}&\makecell{0.77}&\makecell{\textit{0.74}}&\makecell{0.77}&\makecell{0.83}&
    &\makecell{\tb{\textit{*0.83}}}&\makecell{\textit{0.77}}&\makecell{0.77}&\makecell{0.76}&\makecell{0.80}\\
    %\hline
    \makecell{$\alpha_{A}$} & \makecell{0.28}&\makecell{0.15}&\makecell{0.10}&\makecell{0.17}&\makecell{0.23}&
    &\makecell{0.25}&\makecell{0.14}&\makecell{0.09}&\makecell{0.16}&\makecell{0.23}\\
    %\hline 
    \makecell{$\alpha_{E}$} & \makecell{0.24}&\makecell{0.16}&\makecell{0.14}&\makecell{0.14}&\makecell{0.24}&
    &\makecell{0.21}&\makecell{0.15}&\makecell{0.13}&\makecell{0.13}&\makecell{0.18}\\
    %\hline 
    \makecell{$\alpha'_{A}$} &\makecell{0.01}&\makecell{\textbf{0.06}}&\makecell{\textbf{0.08}}&\makecell{0.03}&\makecell{0.03}&
    &\makecell{0.02}&\makecell{0.06}&\makecell{0.08}&\makecell{0.02}&\makecell{\textbf{0.03}}\\ 
    %\hline
    \makecell{$\alpha'_{E}$} &\makecell{\textbf{0.02}}&\makecell{0.02}&\makecell{\textbf{0.03}}&\makecell{0.05}&\makecell{0.03}&
    &\makecell{0.02}&\makecell{0.03}&\makecell{0.03}&\makecell{0.05}&\makecell{\textbf{0.03}}\\ 
    %\hline
    \makecell{$\mathcal{E}$ dist.} & \makecell{0.27}&\makecell{0.29}&\makecell{0.29}&\makecell{0.28}&\makecell{0.28}&
    &\makecell{\textbf{*0.29}}&\makecell{\textbf{0.30}}&\makecell{\textbf{0.30}}&\makecell{\textbf{0.29}}&\makecell{0.29}\\
    \bottomrule
  \end{tabular}
  \caption{Experimental results across two static affordance and temporal \textit{time of day} user classes. Values in bold represent the top value, while marked with * are results where the difference is statistically
significant (t-test with a $0.05$ criterion).}.
  \label{tab:recap}
\end{table*}%
}
%
%
%
%{combined} effect of temporal time-of-day preference upon affordance adoption %behaviours and subsequent reactions to recommendations 
%
%{For each affordance class \tb{$a$,\oe,$o$,$o+$}, we examine whether each %behavioural measure differs significantly in a given time-of-day class with %respect to the other time-of-day classes.} 
%
We next consider the extent to which a given affordance class varies behaviourally alongside temporal preference for platform usage. For instance, does an $o+$ user who has preference for daily usage behave the same as one who has preference for nightly activity?
For each affordance class, we examine whether behavioural measure marked as a top-value respective to their counter-parts in remaining temporal classes are significantly greater. We perform a single-tail Welch’s unequal-variance paired t-test \cite{ttest}. Table \ref{tab:recap} marks a value as significant if  $p > \alpha/3$ (i.e. after adjusting for errors via Bonferroni correction to control for Type I errors) for all remaining temporal counterparts. For instance, considering $(o+, |P|)$, we check if we observe a greater marked or significant difference between all rounders and all other time-of-day classes.
Results are shown in Table \ref{tab:recap}. Largely, we observe time-of-day classes to have no significant effect upon distorting behavioural dynamics for each affordance class. However, one outlier remains, the \textit{all rounder}. Remarkably, this class appears to display distinct behavioural dynamics for almost all affordance classes bar \oe, divulging in high levels of platform usage but at the cost of saturating listening catalogue reflected in the high average $|P|$  and $R$ levels (respectively $27.32K$ and $0.88$). 
\newline\noindent Such findings suggest time-of-day preferences can have a significant effect in mediating surface level activity amongst affordance classes but with no clear downstream propagation to a user's deeper musical preference in terms of varied audio content consumed and preference for item adoption -- a theory which we shall now test empirically. 

\smallskip
\noindent\textbf{Disentangling heterogeneous platform behaviour.} 
To disentangle the influence of time-of-day preference and affordance adoption on user item adoption and reactions to recommendations we next perform a factorial ANOVA, shifting each behavioural attribute to be the dependent variable whose variance we seek to explain. We primarily fit our data to an OLS  model $Y = \beta_0 + \beta_1F + \beta_2T + \beta_3FT + \varepsilon$ (where $F$ and $T$ represent affordance and time-of-day labels respectively) before subsequently applying a factorial ANOVA. Due to space constraints, the full ANOVA results table is not included however we now detail the most relevant results to this work.
As hypothesised, we observe the only effect size ($\eta_p^2$) for which time-of-day classes may have a both moderate and significant effect is with regard to a user's activity $|P|$. On the contrary, affordance classes offer a moderate-to-high explanatory factor for the variance of the remaining behavioural attributes, {foremost adoption}. Perhaps most interestingly, the effect of affordance classes on $\alpha_A$ is particularly strong ($0.11$) implying that a user's decision to adopt items into their organic catalogue may, as hypothesised, be principally a product of adopting recommendation affordances. %Perhaps most interestingly, the effect of affordance classes on \tb{$\alpha_A$} is particularly strong ($0.11$) implying that a user's decision to adopt may, as hypothesised, be principally a product of affordance adoption. %This is also reflected in high negative correlation between \tb{$|P_A|/|P|$ and $\alpha_A$} ($r=0.34, p=8.09e-74$).

For completeness we ultimately examine the effect of sequential time-of-day and affordance adoption influence on the notion of what is meant by an organic stream. 
Contrary to our preliminary belief that organic access acted as a benchmark for Deezer platform exploration we observe users to actually be more algorithmic and editorial than first thought, albeit indirectly. Considering a stream to belong to the affordance in which it was adopted as opposed to organic we recompute centroids for each affordance adoption class. In cases where a stream was both adopted via $A$ and $E$ we deem that the item was adopted via the affordance which had the most streams prior to adoption. 
As shown in Fig.~\ref{fig:affordClust}, we observe all affordance adoption classes centroids to experience a marked shift towards $A$ and $E$ poles -- even more so for users who are already closer to these poles i.e., particularly for $a$ and \oe users.  
\section{Concluding remarks}\label{sec:discussion}
In a time where the modern music streaming platform encapsulates a myriad of modes of accessing content, users may and do personalise their platform use in highly varied ways. % is, as we have shown, more abundant than ever.
%In a time where the modern music streaming platform encapsulates a myriad of modes of accessing content, the potential for users to personalise their platform use in highly varied ways is, as we have shown, more abundant than ever.
%
By acknowledging, assuming and confirming the diversity of user platform behaviour, our work traces the interconnected yet surprisingly sequential factors which drive affordance and item adoption. Our results paint a highly complex picture of user platform behaviour whereby time-of-day preference mediates low-level platform behaviour (activity levels) and affordance adoption distributions, while affordance adoption preference mediates the ultimate higher-level decision to adopt content into one's $O$ catalogue, a factor which is indeed more reflective of musical taste. 

Coming full circle, the heterogeneity of item adoption and its subsequent impact in one's organic catalogue brings into question the nature of what constitutes an organic stream - after taking into consideration the role of adoption, users are indeed found to be markedly less organic than was initially thought. 
This, in turn, may redefine what affordance adoption really is. 
Although beyond the scope of this work, we also suggest that a fruitful direction for future work would be to appraise item adoption relative to affordance adoption, first and foremost by differentiating the impact of repeated exposure along user behavioural classes or by estimating the bidirectional transfer of items between $A$ and $E$ affordances. What is more, we also recommend to explore the role of temporal preference at varied degrees of abstraction be it weekly, monthly or longitudinal.

On the whole, this work aims to hint at a direction that currently remains relatively unexplored in the scholarship concerning the impact of RS on the diversity of user consumption: that user behaviour determines how recommendation affordances are being adopted and apprehended. %, rather than the other way around.
In practice, this type of work and approach could be utilised at the platform level to further the development of context-dependent RS%knowing users use streaming with highly varied behaviours in turn
, providing musical recommendations which are far more suited to the high variety of user's driving use cases.

\clearpage
% For bibtex users:
\bibliography{ISMIR2021_template}
\end{document}